# Effective penetration length and interstitial vortex pinning in superconducting films with regular arrays of defects

J. del Valle,[1,b)] A. Gomez,[1,c)] E. M. Gonzalez,[1,2] and J. L. Vicent[1,2, a)]

[1]*Departamento de Física de Materiales, Facultad de Ciencias Físicas, Universidad Complutense,*

*28040 Madrid, Spain*

[2]*IMDEA-Nanociencia, c/ Faraday 8, Cantoblanco, 28049 Madrid, Spain*

*Abstract.* In order to compare magnetic and non-magnetic pinning we have nanostructured two superconducting films with regular arrays of pinning centers: Cu (non-magnetic) dots in one case, and Py (magnetic) dots in the other. For low applied magnetic fields, when all the vortices are pinned in the artificial inclusions, magnetic dots prove to be better pinning centers, as has been generally accepted. Unexpectedly, when the magnetic field is increased and interstitial vortices appear, the results are very different: we show how the stray field generated by the magnetic dots can produce an effective reduction of the penetration length. This results in strong consequences in the transport properties, which, depending on the dot separation, can lead to an enhancement or worsening of the transport characteristics. Therefore, the election of the magnetic or non-magnetic character of the pinning sites for an effective reduction of dissipation will depend on the range of the applied magnetic field.

Vortex dynamics in superconducting thin films have been subject of intense research during last years (1-3). Lithography techniques have allowed to nanostructure these films with sizes similar to the superconducting characteristic lengths, deeply affecting vortex pinning potentials and, hence, transport properties of the superconductor. As a result, a variety of new phenomena has been observed, such as commensurability,[1] ratchet[2] or guided motion effects (4). The effect of mesoscopic magnetic inclusions on the superconducting properties has been thoroughly studied (5). Interplay between magnetism and superconductivity has led to the appearance of novel behaviors, such as field induced superconductivity (6), with promising applications as memory or sensing devices (7). Despite all this progress, the fundamental problem of which is the best way to reduce dissipation in superconductors is still an open subject of great technological interest.

___________________________

[a)] Author to whom correspondence should be addressed. Electronic mail: jlvicent@ucm.es
[b)] Present address: Department of Physics, University of California-San Diego, CA 92093, USA.
[c)] Present address: Centro de Astrobiología, INTA-CSIC, 28850 Torrejón de Ardoz, Spain.

Recently, some authors have addressed the problem of the optimum geometry, finding that a graded pinning landscape provides the best transport properties over a broad range of magnetic fields (8,9). Yet, it is still uncertain which kind of materials should be used to induce effective pinning landscapes. Most authors have found magnetic inclusions to provide stronger individual vortex pinning than non-magnetic ones (10-12). However, this does not necessarily mean less dissipation (13), since, at high applied magnetic fields, most vortices occupy interstitial positions and are not directly pinned by the artificial pinning centers, but by elastic strains of the vortex lattice. In this case, onset of dissipation might be determined by the elastic softening of the vortex lattice. How the magnetism of the inclusions can affect this softening is a problem which had not been addressed yet, despite it is the key to achieve better transport properties in a broad field range. In this letter, we compare the magnetotransport properties of two nanostructured Nb thin films: one with an array of non-magnetic Cu dots and the other with an identical array of magnetic Permalloy (Py) dots. While the Py dots sample shows better pinning properties at low fields, we show that stray fields can produce an effective decrease of the penetration length, profoundly affecting vortex dynamics at higher fields: this can enhance or worsen transport properties over an order of magnitude, depending on the dot spacing. These results prove the great impact of the magnetism on the interstitial vortex dynamics, and reveal that the optimum pinning geometry is very dependent on the magnetic character of the inclusions.

Both samples are based on a rectangular (a = 400 nm, b = 600 nm) array of nanodots (200 nm diameter and 40 nm thickness) fabricated using electron beam lithography and magnetron sputtering (Cu in one case and Py in the other). A 100 nm superconducting Nb thin film Nb ($T_C$ = 8.5 K) was grown on top of these arrays using magnetron sputtering in a chamber with base pressure $5 \cdot 10^{-8}$ Torr. In order to test the interstitial vortex dynamics when pushed along the two different directions of the array, optical lithography and reactive ion etching were used to define a cross shaped, 40 μm wide, measurement bridge. Magnetotransport measurements have been carried out in a commercial Helium cryostat, with a 9 T superconducting magnet and temperature stability better than ±2 mK. The stray field generated by the Py dots has been calculated using OOMMF micromagnetic simulations. Due to Py low anisotropy, magnetization is arranged in a magnetic vortex (14) configuration with a small core (less than 10 nm), yielding a linear and reversible out of plane hysteresis loop around H=0, with no out of plane remanent magnetization, $M_{R,z}$ = 0. This is desirable, as it is well known that nanomagnets with $M_{R,z} \neq 0$ often lead to shifts of $T_C$ (6,7,15), which would obscure a direct comparison with the results obtained in the Cu dots sample.

Figure 1 shows critical currents of both samples as function of the temperature at the first matching field, when there is one vortex trapped in each dot ($H_m = \phi_0/(a \cdot b) = 87$ Oe). In this situation, the critical current is proportional to the pinning force, which is the same on each vortex. As observed in previous studies(10-12), magnetic dots work better as pinning centers, with critical currents considerably higher in the whole temperature range. As is well known, vortex cores are surrounded by supercurrents that concentrate (screen) the magnetic field within (further than) $\lambda$, the penetration length of the superconductor. The stray field of a magnetized dot does precisely this effect: concentrating the flux inside the dot while screening it on the outside. For this reason, supercurrents will be weaker if a vortex is trapped on a magnetic dot that if it is trapped in a non-magnetic one, resulting in lower vortex line energy and, thus, a stronger pinning potential (16). Inset in figure 1 displays the extra flux (generated by the stray field) that threads each Py dot, as a function of the external applied field. As can be noted, even for low fields, this flux is not negligible and is comparable to $\lambda$, producing a substantial reduction of the supercurrents that explains the difference in pinning strengths between both samples. According to the figure, this difference is expected to increase with the applied field.

Figure 2 shows the resistance as a function of the magnetic field for both samples. Py dot sample shows much lower resistance for low applied fields due to the higher pinning strength. Surprisingly, as the magnetic field is increased, a crossover takes place, with the Cu dots sample showing less resistance. This happens despite the fact that magnetic pinning is expected to increase its intensity with the magnetic field. Inset in figure 2 shows the critical temperature of both samples as a function of the magnetic field: both display the same behavior. This situation rules out the possibility that the increasing stray fields, generated by the Py dot, reduce the critical temperature of the film, leading to the observed crossover. The origin of this unexpected behavior must be on the vortex dynamics. The number n of vortices that can fit on a pinning site can be estimated by $n = D/4\xi(T)$, where D is the diameter of the dot and $\xi(T)$ the coherence length (17). In our case, in the temperature range in which electrical measurements can be properly performed (above 0.9 Tc), only one vortex fits on each dot, meaning that in the crossover region transport properties will be dominated by the interstitial vortices dynamics. As already pointed out, these vortices are not directly pinned, and elastic strains avoid their motion when a Lorentz force is applied. These strains are a result of the repulsive force acting between neighboring vortices:

$$F_{1\rightarrow 2} = \frac{\phi_0}{4\pi} B_1(r) = \frac{\phi_0^2}{8\pi^2\lambda^2} K_0\left(\frac{r}{\lambda}\right) \qquad (1)$$

Where $B_1(r)$ is the magnetic field profile created by a vortex, and $K_0$ the zeroth-order Hankel function, which limits the range of the interaction to distances in the order of λ.

As shown in inset of figure 1, increasing the magnetic field rapidly increases the flux generated by stray fields, condensing the magnetic field inside the Py dot and effectively reducing the penetration length of the vortex. This vortex size shrinking has been predicted for magnetic superconductors with an intrinsic permeability (18), and has also been observed in Ginzburg-Landau simulations of hybrid superconducting/magnetic systems similar to this one (19). The effective λ reduction yields important consequences in the vortex repulsive interaction, strengthening it for short distances but reducing its range. In general, the interaction becomes stronger for r < λ, and weaker for r > λ.

Using the dirty limit approximation (20) and taking into account $\xi_0 = 38$ nm for pure Nb (21), data can be fitted to obtain the Ginzburg-Landau coherence length, ξ(0) = 9.8 nm; the mean free path $l = 3.5$ nm. Therefore, the penetration length of the Nb films can be estimated:

$$\lambda(T) = 0.715 \frac{\lambda_L(0)}{l} \frac{\xi(0)}{\sqrt{1 - T/T_C}} \qquad (2)$$

Where $\lambda_L(0) = 48$ nm is the London penetration length (22).

Figure 3 shows the critical current in both samples as a function of the magnetic field when vortices are pushed along the two symmetry directions of the array, for a temperature T = 0.97 $T_C$, with λ = 275 nm. As can be seen, the comparative field dependence is absolutely different depending on the direction. It has been observed, both experimentally (23) and in simulations (24), that after depinning, interstitial vortices will start moving along the empty channels defined by the nanodot array, in a plastic motion between rows of trapped vortices. The anisotropic shape of the nanodot array defines two channels as depicted in the insets: a wide one in which the interstitials can flow far from the trapped vortices, keeping always a distance above λ; and a narrow one, in which they are forced to move closer to them, at distances under λ. The effective λ reduction affects both situations very differently. In the first case, the repulsive interaction will become weaker, decreasing elastic strains along that direction. In that situation, the higher dot magnetization, the easier interstitial vortices can depin and start moving. This explains the crossover observed in the magnetoresistance measurements in Fig. 3a. In the second case, repulsive interaction will become stronger, increasing elastic strains. This leads to the opposite situation, as showed in Fig. 3b and the critical current will increase with dot magnetization.

In summary, the use of magnetic nanostructures does not necessarily improve the transport properties of superconducting films. The stray fields created by these structures make them more efficient pinning centers than non-magnetic ones. However, magnetization can lead to an effective reduction of the penetration length, deeply affecting the dynamics of interstitial vortices. It has been found that this could enhance or depress transport properties at high fields, depending on the periodicity of the array, d. For separations $d > \lambda$, the vortex lattice becomes softer, interstitial vortices can move easier and the inclusion of magnetic nanostructures will negatively affect the electric performance. On the contrary, in more packed arrays, $d < \lambda$, magnetic character of the nanostructures considerably enhances transport properties of the sample.

*Acknowledgments:* We thank support from Spanish Ministerio de Economía y Competitividad grant FIS2013-45469 and CAM grant P2013/MIT-2850 and EU COST Action MP1201.

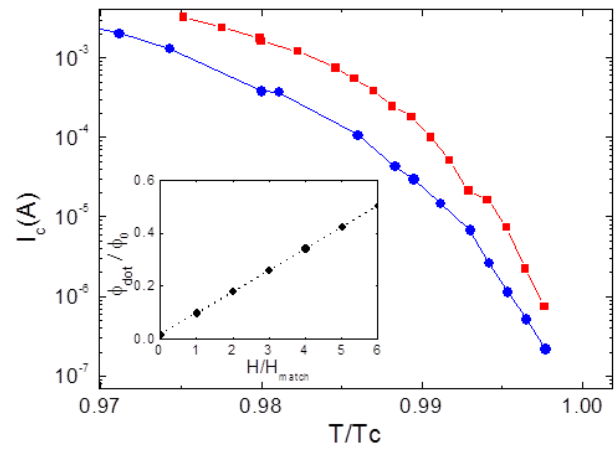

FIG. 1. Critical current as a function of the temperature for Cu (blue dots) and Py (red squares) nanodot samples, at the first matching field $H_m = 87$ Oe. Critical current criterion is 5 μV/mm. Inset: Stray field flux (normalized to $\phi_0 = 2.07 \cdot 10^{-15}$ Wb) threading a Py dot as a function of the applied magnetic field (normalized to $H_m$).

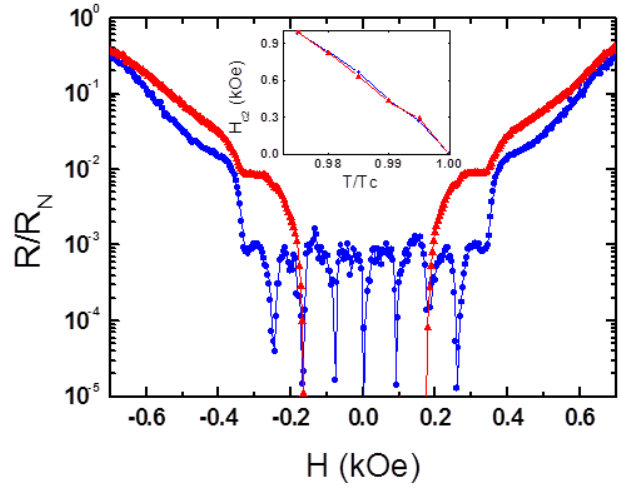

FIG. 2. Normalized resistance to its normal value as a function of the magnetic field for the Cu (blue dots) and Py (red triangles) nanodot sample, measured with 2.5 mA at 0.975 $T_C$, with vortices pushed along the short side of the array. Inset: $H_{C2}$ as a function of $T/T_C$ for the Cu (blue dots) and Py (red triangles) nanodot sample.

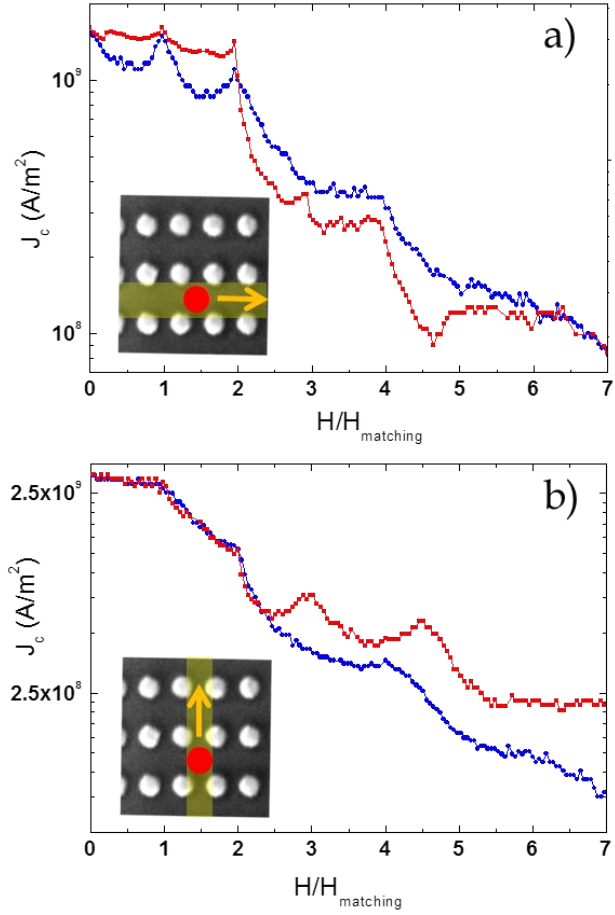

FIG. 3. Critical current as a function of the magnetic field for the Cu (blue dots) and Py (red squares) nanodot sample, measured at 0.975 $T_C$, with vortices pushed along the (a) short side of the array and (b) the long side of the array. Critical current criterion is 5 µV/mm. Inset: Schematic description of the interstitial vortex (in red) motion (orange arrow) along the channels defined by the array (faded yellow).